%% file: main.tex
\pdfoutput=1
\documentclass[
	epjc3,
	twocolumn
]{svjour3}
\usepackage[pdftex]{graphicx}    

\usepackage[utf8]{inputenc}
\usepackage[english]{babel}
\usepackage{textcomp}
\usepackage{mathptmx}

\usepackage{subscript}
\usepackage{microtype}

\usepackage{amsmath}
\usepackage{amssymb}
\usepackage[separate-uncertainty=true,
 range-units = single,
 range-phrase =\:\textendash\:
]{siunitx}

\usepackage[
	backend=bibtex,
style=numeric-comp,
sorting=none,
eprint=false,
firstinits=true,				
maxnames=3,
]{biblatex} 
\renewbibmacro*{in:}{} 

\bibliography{Bibliography}

\usepackage{hyperref}
\hypersetup{
    colorlinks=true,
    linkcolor=black,
    citecolor=black,
    filecolor=black,
    urlcolor=black,
}

\newcommand{\EeV}{\exa\electronvolt}
\DeclareSIUnit\parsec{pc}
\DeclareSIUnit\lightyear{ly}
\DeclareSIUnit\gauss{G}
\DeclareSIUnit\Sigma{$\sigma$}
\DeclareSIUnit\year{yr}
\DeclareSIUnit\years{yr}
\DeclareSIUnit\erg{erg}

\title{\boldmath
Search for patterns by combining cosmic-ray energy and arrival directions at
the Pierre Auger Observatory
}
\author{The Pierre Auger Collaboration\thanksref{augermail, augeradress}}

\thankstext{augermail}{
	email: auger\_spokespersons@fnal.gov
} 
\institute{The Pierre Auger Observatory, Av. San Mart\'{\i}n Norte 306, 5613 Malarg\"{u}e, Mendoza, Argentina; \url{http://www.auger.org}
	\label{augeradress}}
	\date{Published in the European Physical Journal C as doi:10.1140/epjc/s10052-015-3471-0}
\begin{document}\sloppy
\maketitle
\begin{abstract}
	Energy-dependent patterns in the arrival directions of cosmic rays are
	searched for using data of the Pierre Auger Observatory.  We investigate
	local regions around the highest-energy cosmic rays with $E \geq
	\SI{6d19}{\electronvolt}$ by analyzing cosmic rays with energies above $E =
	\SI{5d18}{\electronvolt}$ arriving within an angular separation of approximately \SI{15}{\degree}.  We
	characterize the energy distributions inside these regions by two independent
	methods, one searching for angular dependence of energy-energy correlations
	and one searching for collimation of energy along the local system of
	principal axes of the energy distribution.  No significant patterns are
	found with this analysis. The comparison of these measurements with
	astrophysical scenarios can therefore be used to obtain constraints on
	related model parameters such as strength of cosmic-ray deflection and
	density of point sources.
\end{abstract}

\input{Introduction}

\input{Methods}

\input{ToyMonteCarlo}

\input{Results}
\input{Discussion}

\input{Conclusions}
\input{acknowledgments}

\printbibliography[title=References]

\input{author_list_latex.tex}
\end{document}

%% file: Introduction.tex
\section{Introduction}
The long-standing question about the origin and nature of the ultra-high
energy cosmic rays (UHECRs) is yet unanswered. 
Presumably, UHECRs are charged nuclei of extragalactic origin. They are
deflected in extragalactic magnetic fields and the magnetic field of the Milky
Way such that their arrival directions may not point back to their
sources~\cite{Kotera2011}.  The structure, strength, and origin of these cosmic
magnetic fields are open questions in astrophysics as
well~\cite{Ryu2012, Widrow2012}.  Consequently, UHECRs can also be considered to be
probes of the magnetic fields they traverse~\cite{Lee1995, Lemoine1997} as the
deflections lead to energy-dependent patterns in their arrival directions,
and an analysis of such patterns may allow for conclusions on the strength and
structure of the fields.

The Pierre Auger Observatory~\cite{PAO2004, PAO2010b} is currently the largest
experiment dedicated to observations of UHECRs.  In
2007, we reported evidence for a correlation of events with energies above
\SI{60}{\EeV} ($\SI{1}{\EeV} = \SI{d18}{\electronvolt}$) with the distribution of nearby extragalactic
matter~\cite{PAO2007, PAO2008}. An update of the analysis yielded a
correlation strength which is reduced compared
to the initial result~\cite{PAO2010d}.  Further searches for anisotropy using
variants of autocorrelation functions~\cite{PAO2012b}  yielded no
statistically-significant deviation from isotropic scenarios.  Following this
observation, constraints on the density of point sources and magnetic fields
have been reported~\cite{Abreu2013a}.  Also a direct search for
magnetically-induced alignment  in the arrival directions of cosmic rays
assuming they were protons has been performed without uncovering so-called
multiplet structures beyond isotropic expectations~\cite{PAO2012} .

Nevertheless, if the highest-energy cosmic rays with $E > \SI{60}{\EeV}$ are
tracers of their sources and even if their deflection in magnetic fields is
dependent on their nuclear charges, some of the lower-energy cosmic rays in a
region around them may be of the same origin. From deflections both in
extragalactic magnetic fields and the magnetic field of the Milky Way, their
distribution of arrival directions may show energy-dependent patterns. In
particular a circular `blurring' of the sources is expected from deflection in
turbulent magnetic fields, while energy dependent linear structures are
expected from deflection in coherent magnetic fields.

In this report, we investigate the local regions around cosmic rays 
with  $E \geq \SI{60}{\EeV}$ by analyzing cosmic rays with energies above $E = 5$~EeV 
arriving within an angular separation of \SI{0.25}{\radian}. The lower energy cut 
just above the ankle is motivated by the assumption that the selected 
cosmic rays are predominantly of extragalactic origin. The angular 
separation cut has been optimized from simulation studies and will be 
explained below.

We use two methods to characterize the energy distributions inside the local
regions. In one method we study energy-energy correlations between pairs of
cosmic rays depending on their angular separation from the center of the region.
With this measurement we search for signal patterns 
expected from particle deflection in turbulent magnetic fields.
In the second method we decompose the directional energy
distribution of the cosmic rays along its principal axes.  This general decomposition method
imposes no requirement on the sign of the cosmic-ray charge, or the charge
itself. Beyond measuring the strength of collimation along principal axes, the
axis directions of the individual regions around the highest-energy cosmic rays
potentially reveal local deflection patterns due to magnetic fields.

Both methods were originally studied in particle physics, 
and were referred to as energy-energy correlations and thrust observables,
respectively~\cite{Basham1978, Brandt1964}. Simulations of their application in cosmic-ray physics 
have demonstrated the capability to reveal effects from coherent and turbulent 
magnetic fields~\cite{Erdmann2009, Erdmann2013}.

This paper is structured as follows.  The observables of the energy-energy
correlations and the principal-axis analysis are defined in
Section~\ref{sec:Methods}.  Their response to structure potentially expected from
deflection in magnetic fields is illustrated using a simplified model in Section~\ref{sec:ToyMC}.  The
measured distributions of the observables using data of the surface detector of
the Pierre Auger Observatory are presented in Section~\ref{sec:Results}. In
Section~\ref{sec:Discussion}, we first analyze the directional characteristics
of the measured principal axes by studying their reproducibility.  We then
present a comparison of the measurements with an astrophysical model of UHECR
origin and propagation, and determine constraints on the source density, and
the strength of cosmic-ray deflection as the two dominant model parameters.

%% file: Methods.tex
\section{Definitions}
\label{sec:Methods}
In this section we introduce the main components used for the 
measurement. We first define the local regions in which we analyze the 
cosmic-ray energies and arrival directions. We then explain the 
energy-energy correlation observable and its angular dependence. 
Finally, we present the method of calculating the principal axes of the 
energy distribution which results in the three values to characterize 
the strength of collimation along each axis, and the directions of the 
axes themselves.

\subsection{Region of Interest}
The observables used here are calculated from the events detected in a bounded
region in the sky, here denoted as `region of interest' (ROI). To minimize the
statistical penalty from multiple tries, we do not scan the entire sky but
investigate a limited number of ROIs located around events with an energy above
\SI{60}{\EeV}. This energy cut is  motivated by the limitation of the
propagation distance by, e.g., the GZK effect~\cite{Greisen1966, Zatsepin1966}
and corresponds to the energy used in the AGN correlation
analysis~\cite{PAO2007}. The size of the ROIs, i.e. the maximum angular
separation of a UHECR belonging to the ROI to the center of the ROI, is set to
\SI{0.25}{\radian}.  To choose these values we simulated the UHECR propagation
in magnetic fields with the UHECR simulation tool PARSEC~\cite{Bretz2013} for
different strengths of the deflection and source density. The simulations were
analyzed with varying choices of parameters. The chosen values maximize the
power of the observables to discriminate between scenarios with strong
deflections and isotropic scenarios~\cite{Schiffer2011, Winchen2013}.  To avoid
a possible bias of the characterization of the ROI, we exclude the cosmic ray
seeding the ROI  from the calculation of the observables.

\input{EEC.tex}
\input{Thrust.tex}

%% file: EEC.tex
\subsection{Energy-Energy Correlations}
Energy-energy correlations (EECs) are used to obtain information on the
turbulent part of galactic and extragalactic magnetic
fields~\cite{Erdmann2009}. The concept of the EEC was originally developed  for
tests of quantum chromodynamics (QCD)~\cite{Basham1978}. The
Energy-energy correlation $\Omega_{ij}$ is calculated for every pair of UHECRs
$i,j$ within a ROI using
\begin{equation}\label{eq:EEC}
 \Omega _{ij}= \frac{(E_i-\langle E(\alpha_i) \rangle)\,(E_j-\langle E(\alpha_j) \rangle) }{E_i \, E_j}.
\end{equation}
Here $E_i$ is the energy of the UHECR $i$ with the angular separation 
$\alpha_i$ to the center of the ROI. $\langle E_i(\alpha_i) \rangle$ is
the average energy of all UHECRs at the angular separation $\alpha_i$ from
the center of the ROI.

The cosmic rays in a ROI can be separated into a signal fraction, whose arrival
direction is correlated with the energy, and an isotropic background fraction.
The values of $\Omega_{ij}$ can be positive or negative depending on the
cosmic-ray pair having energies above or below the average energies. An angular
ordering is measured in the following sense. A pair of cosmic rays, one being
above and the other below the corresponding average energy, results in a
negative correlation $\Omega_{ij} < 0$. This is a typical case for a background
contribution. A pair with both cosmic rays having energies above or below the
average energy at their corresponding angular separation gives a positive
correlation $\Omega_{ij} > 0$.  Here both signal and background pairs are
expected to contribute. As the correlations are determined as a function of the
opening angle to the center of the ROI, circular patterns can be found that are
expected from turbulent magnetic deflections which are sometimes viewed as
random-walk propagation.

We present the angular distribution of the EEC as the average 
distribution of all ROIs. Each value $\Omega_{ij}$ is taken into account twice, 
once at the angular separation $\alpha_i$ and once at $\alpha_j$. 

%% file: Thrust.tex
\subsection{Principal Axes}
To further characterize energy-dependent patterns within each individual ROI, we 
calculate the three principal axes of the energy distribution which we 
denote as $\vec{n}_{k=1,2,3}$. For this we successively maximize the quantity 
\begin{equation}
	T_k = \max_{\vec{n}_k} \left(\frac{\sum_i |\omega_i^{-1}\; \vec{p}_{i}\cdot
	\vec{n}_k|}{\sum_i |\omega_i^{-1}\; \vec{p}_i|} \right)
	\label{eq:ThrustValues}
\end{equation}
with respect to the axes $\vec{n}_{k}$ starting with $k=1$.
Here  $\vec{p}_i$ is the cosmic-ray momentum and $\omega_i$ the corresponding exposure of
the detector~\cite{Sommers2001} in the direction of particle $i$. The values of
$T_{k=1,2,3}$ quantify the strength of the collimation of the particle momenta
along each of the three axes $\vec{n}_{k=1,2,3}$ of the principal system.  We
denote $T_{k=1,2,3}$  as thrust observables following previous studies of
perturbative QCD in particle collisions~\cite{Brandt1964, Farhi1977}.

For $k = 1$ the quantity $T_1$ is called the `thrust' and consequently the
first axis of the principal system $\vec{n}_1$ is called `thrust axis'. For the
second axis the additional condition $\vec{n}_1 \perp \vec{n}_2$ is used in
Equation~\eqref{eq:ThrustValues}. The resulting value $T_2$ is denoted as `thrust
major', the axis as `thrust-major axis'. Finally, the third quantity $T_3$ is
called `thrust minor' with corresponding `thrust-minor axis'.  For the
thrust-minor axis $\vec{n}_3$ it is $\vec{n}_1 \perp \vec{n}_2 \perp \vec{n}_3$
which renders the maximization in Equation~\eqref{eq:ThrustValues} trivial. From
this definition follows $T_1 > T_2 > T_3$.

In arbitrarily defined spherical coordinates $(r, \phi, \theta)$ with
orthonormal basis $(\vec{e}_r, \vec{e}_\phi, \vec{e}_\theta )$ and the
observer at the center, the momenta of the
particles at the high energies  considered here can be written as $\vec{p}_i =
\vert E_i \vert \vec{e}_{r_i}$ with the energy $E_i$  and the radial
unit vector $\vec{e}_{r_i}$ in the arrival direction of particle $i$.   The
thrust axis is thus the radial unit vector $\vec{e}_r$ pointing to the local
barycenter of the energy distribution, and the thrust value is a measure for the
energy-weighted strength of clustering of the events. 
For no dispersion of the particles in the
region it takes on the value $T_1 = 1$, whereas for an isotropic distribution in a circular
region the expectation value of $T_1$ depends dominantly on the size of the ROI~\cite{Winchen2013}. 

The thrust-major and
thrust-minor axes can consequently be written as 
\begin{eqnarray}
	\vec{n}_{2} = \cos{\xi_{2}} \, \vec{e}_\phi + \sin{\xi_{2}} \, \vec{e}_\theta \\
	\vec{n}_{3} = \cos{\xi_{3}} \, \vec{e}_\phi + \sin{\xi_{3}} \, \vec{e}_\theta
\end{eqnarray}
with the angles $\xi_2$ and $\xi_3 = 90^\circ + \xi_2$ between the corresponding axes
and the vector $\vec{e}_\phi$. Using this together with
Equation~\eqref{eq:ThrustValues}, the thrust-major $T_2$ becomes maximal if $\vec{n}_2$ is
aligned with a linear distribution of UHECR arrival directions. The thrust-major axis thus points
along threadlike structures in the energy distribution of UHECRs.  As the thrust
minor axis is chosen perpendicular to $\vec{n}_1$ and $\vec{n}_2$ it has no
physical meaning beyond its connection to the thrust-major axis. However, the
thrust-minor $T_3$ gives meaningful information as it denotes the
collimation strength perpendicular to the thrust-major axis.

Note that in a perfect isotropic scenario, the energy distribution within the
plane defined by $\vec{n}_2$ and  $\vec{n}_3$ exhibits perfect symmetry. The
values of $T_2$ and $T_3$ are approximately equal, and the axis directions are accidental.
However, even with a small signal contribution beyond an isotropic background,
the circular symmetry in the $(\vec{n}_2 , \vec{n}_3)$ plane is broken giving
rise to unequal values of $T_2$ and $T_3$.  In addition, the direction of the
thrust-major axis then reveals valuable directional information.
 This directional information can be compared to the direction of deflection
 obtained in a multiplet analysis~\cite{PAO2012}. However, in contrast to the multiplet
analysis the principal axes analysis does not require a uniform charge of the
cosmic rays. Its sensitivity is driven by the total deflection amount.

%% file: ToyMonteCarlo.tex
\section{Benchmark Distributions for Coherent and Turbulent Magnetic Fields}
\label{sec:ToyMC}
For obtaining a general understanding of the energy-energy correlations and the
thrust observables, we use simple scenarios of cosmic-ray deflections in
magnetic fields to demonstrate resulting distributions. First we describe the
procedure for simulating mock data representing cosmic-ray deflection in
turbulent and coherent magnetic fields. For different quantitative mixtures of these field
types we then present the distributions of the energy-energy correlations
and finally discuss the resulting thrust distributions.

\subsection{Simulation Procedure}
To demonstrate the sensitivity of the observables to deflections expected from
magnetic fields, we simulate a ROI with UHECRs in a simplified scenario.
The deflection in cosmic magnetic fields is supposed to result in two different
kinds of patterns in the arrival direction of the UHECRs. First, if the UHECR's
trajectory resembles a directed random walk, a symmetric blurring of the source
is expected. Second, if the particles are deflected in
large-scale coherent fields, e.g. in the Milky Way, an energy ordering of the
UHECRs in threadlike multiplets is expected.

\begin{figure*}[tb]
	\begin{center}
		\includegraphics[width=\textwidth]{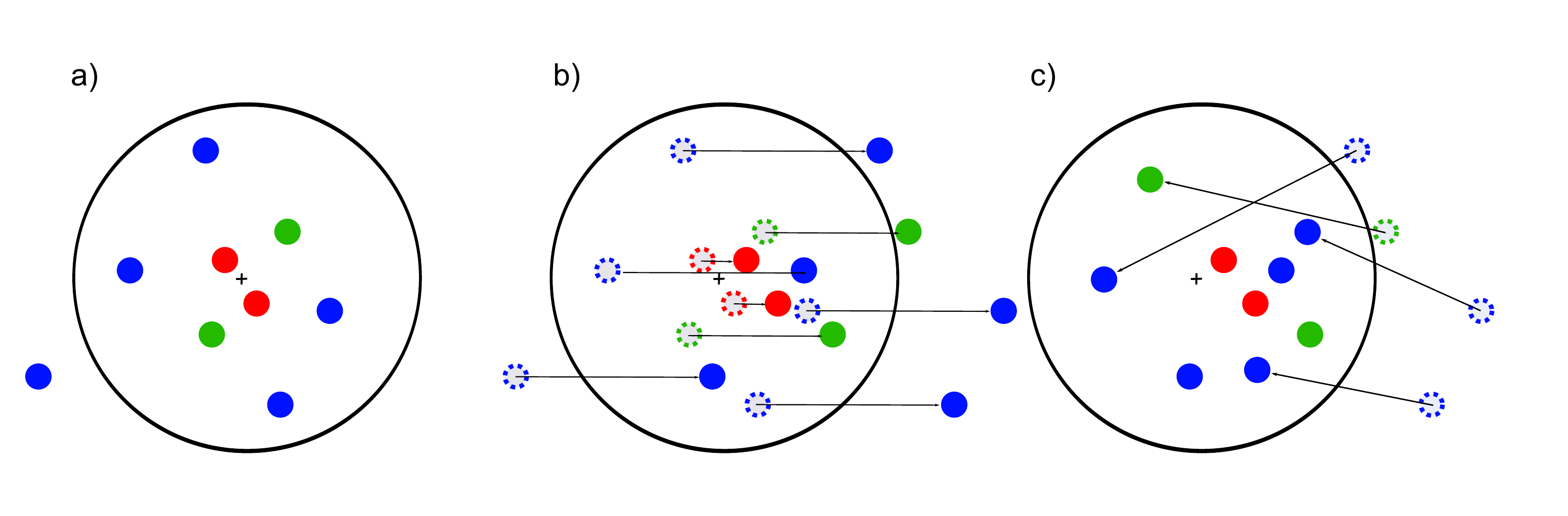}
	\end{center}
	\caption{Generation of anisotropically distributed UHECRs in a region
		of interest. \textbf{(a)} First, UHECRs are
	distributed symmetrically around the center of the ROI using a Fisher
	distribution with energy dependent concentration parameter according
	to Equation~\eqref{eq:CTurbulent}. \textbf{(b)} The UHECRs are  then deflected
	in one direction using Equation~\eqref{eq:CCoherent}. \textbf{(c)} UHECRs
	deflected outside of the ROI are moved to a random position inside the
	region.}
	\label{fig:ToyMCSketch}
\end{figure*}
Here we model the distribution of UHECRs in a region around the source as
a superposition of both effects. Events in this region of interest are
generated in three steps as sketched in Figure~\ref{fig:ToyMCSketch}.
First, the UHECRs are distributed around the center of the ROI 
following  a Fisher
distribution~\cite{Fisher1953} with probability density 
\begin{equation}
	f(\alpha,\kappa) = \frac{\kappa}{4\pi\, \sinh{\kappa}} e^{(\kappa\,
	\cos{\alpha)}}
	\label{FisherDistribution}
\end{equation}
for angle $\alpha$ between cosmic ray and center of the ROI. The Fisher
distribution can be considered here as the normal
distribution on the sphere. The concentration
parameter $\kappa$ is chosen with an energy dependence that emulates the
deflection in turbulent magnetic fields as
\begin{equation}
	\kappa = C_\text{T}^{-2}	E^{2}.
	\label{eq:CTurbulent}
\end{equation}
For small deflections the distribution resembles a Rayleigh distribution
where $\kappa$ is related to the root-mean-square $\delta_{\text{RMS}}$ of the
deflection angles by $\kappa = \delta_{\text{RMS}}^{-2}$ and thus 
\begin{equation}
	\delta_{\text{RMS}} \simeq \frac{C_\text{T}}{E}.
	\label{eq:dRMS_CTRelation}
\end{equation}
A value of $C_\text{T} = \SI{1}{\radian\EeV}$
is equivalent to an RMS of the deflection angle $\delta_{\text{RMS}} =
\SI{5.7}{\degree}$ for 10~EeV particles.  For example, using the usual
parametrization for deflections in turbulent magnetic
fields~\cite{Achterberg1998, Harari2002} this corresponds to the
expected deflection of \SI{10}{\EeV} protons from a source at a distance
$D \approx \SI{16}{\mega\parsec}$ propagating through a turbulent
magnetic field with coherence length $\Lambda \approx
\SI{1}{\mega\parsec}$ and strength $ B \approx \SI{4}{\nano\gauss}$.

Second, a simple model for the deflection in coherent magnetic fields is added
on top of the model for turbulent magnetic fields used above. Here the
individual cosmic rays are deflected in one direction by an angle $\alpha$ that
depends on the energy of the particles according to 
\begin{equation}
	\alpha = C_\text{C}\, E^{-1} \label{eq:CCoherent}
\end{equation}
where the parameter
$C_\text{C}$ is used to model the strength of the coherent deflection.
The procedure is illustrated in Figure~\ref{fig:ToyMCSketch}~(b).

Third, particles deflected outside the region of interest are added as
a background to keep the number of particles in this setup constant
(cf.~Figure~\ref{fig:ToyMCSketch}~(c)).  The energies of all events are chosen
following a broken power law with spectral index $\gamma_1 = -2.7$ below
\SI{40}{\EeV} and $\gamma_2 = -4.2$ above \SI{40}{\EeV} to be comparable with the observed cosmic-ray energy spectrum~\cite{PAO2010a}.

\subsection{Response of the Energy-Energy Correlation}
\begin{figure*}[tbp]
	\includegraphics[width=\textwidth]{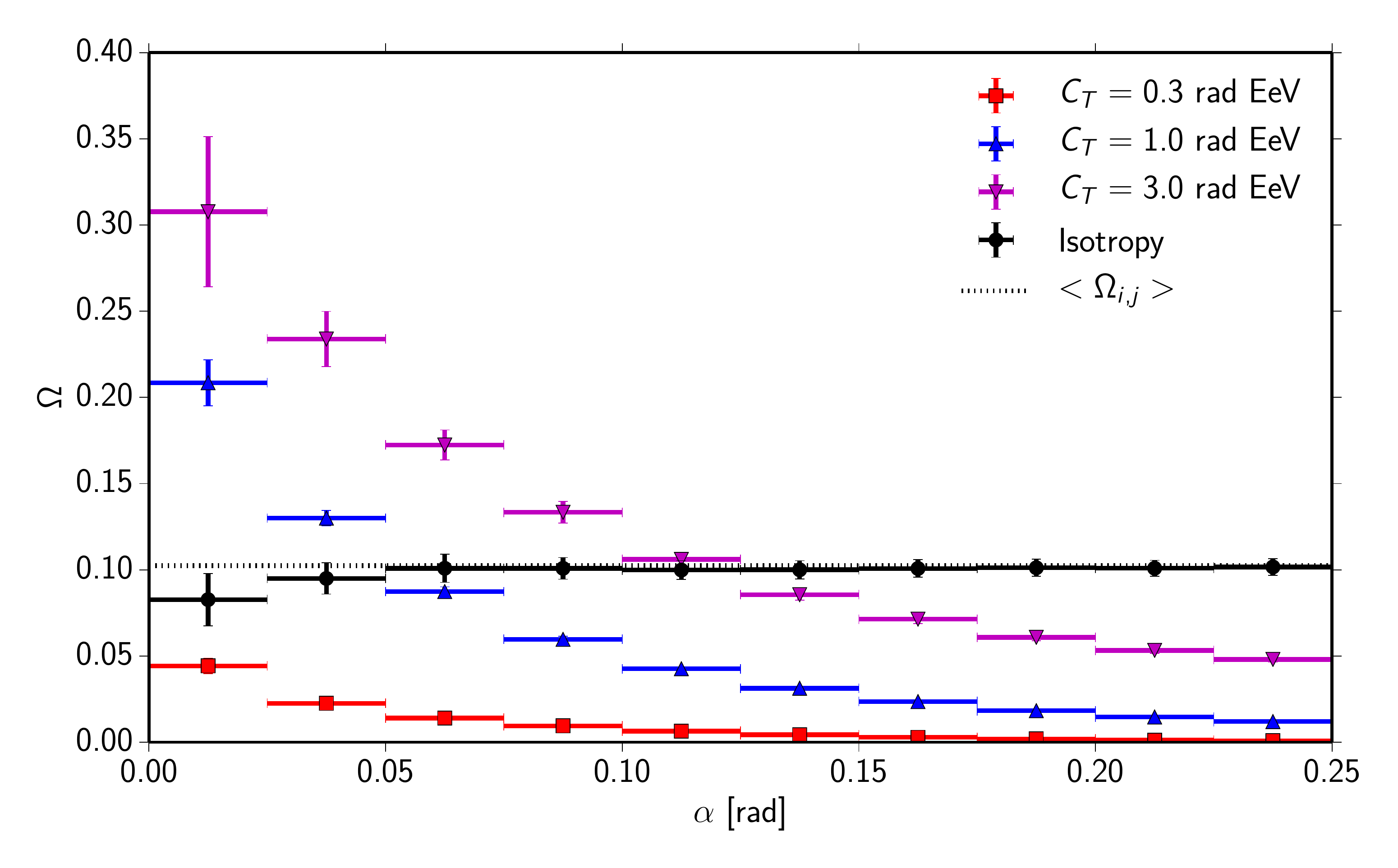}
	\caption{Response of the EEC to typical deflection patterns
	from simulations of three different turbulent deflection strengths with
	$C_\text{T}=\SI{0.3}{\radian\EeV}$ (red squares), $C_\text{T}=\SI{1}{\radian\EeV}$ (blue
	upward triangles) and $C_\text{T}=\SI{3}{\radian\EeV}$ (magenta downward triangles).
	The dashed line marks the isotropic expectation value according to
	Equation~\eqref{eq:expvalue}; black circles denote the result from simulation of
	isotropically distributed UHECRs.}
	\label{fig:EECToyMC}
\end{figure*}
The EEC distributions resulting from simulated scenarios using the three values
for the turbulent deflection strength $C_\text{T} = 0.3, 1.0,\;\SI{3.0}{\radian\EeV}$
are shown in Figure~\ref{fig:EECToyMC}. As the EEC is expected to provide only
minor sensitivity to coherent deflections~\cite{Erdmann2009} $C_\text{C} = 0$ is used
here. For each scenario 50 realizations of an ROI with 300 UHECRs have been
used, which is approximately the number of UHECRs in a low-coverage region of the measurement presented in Section~\ref{sec:Discussion}.
All scenarios are compared with the result for an isotropic distribution
of UHECRs. Without structure in the arrival directions of UHECRs, the EEC
distribution is flat with an expectation value 
\begin{equation}\label{eq:expvalue}
 \left<\Omega_{ij}\right> =\left< \frac{\left( E_i -\left< E \right> \right)\, \left( E_j -\left< E \right> \right)} {E_i\, E_j}\right> = \left(1- \left<E\right> \left<\frac{1}{E}\right> \right)^2.
\end{equation}
For a source signal the typical signature is an increase towards small angles,
as can be seen in Figure~\ref{fig:EECToyMC}. With increasing angular separation 
the UHECRs average energies decrease, and so do the differences between the
UHECR energies and their corresponding average (Equation~\eqref{eq:EEC}).
Consequently, the values of $\Omega_{ij}$ can become small in contrast to a
scenario where all UHECR energies contribute at every angular scale.  The shape
of the EEC distribution in response to a source signal depends on the
deflection pattern. In general it can be seen that a small deflection causes
an increase only in the innermost bins, while a larger deflection will smear
this signature over the whole ROI.

\subsection{Response of the Principal-Axes Analysis}
\begin{figure*}[tbp]
	\includegraphics[width=\textwidth]{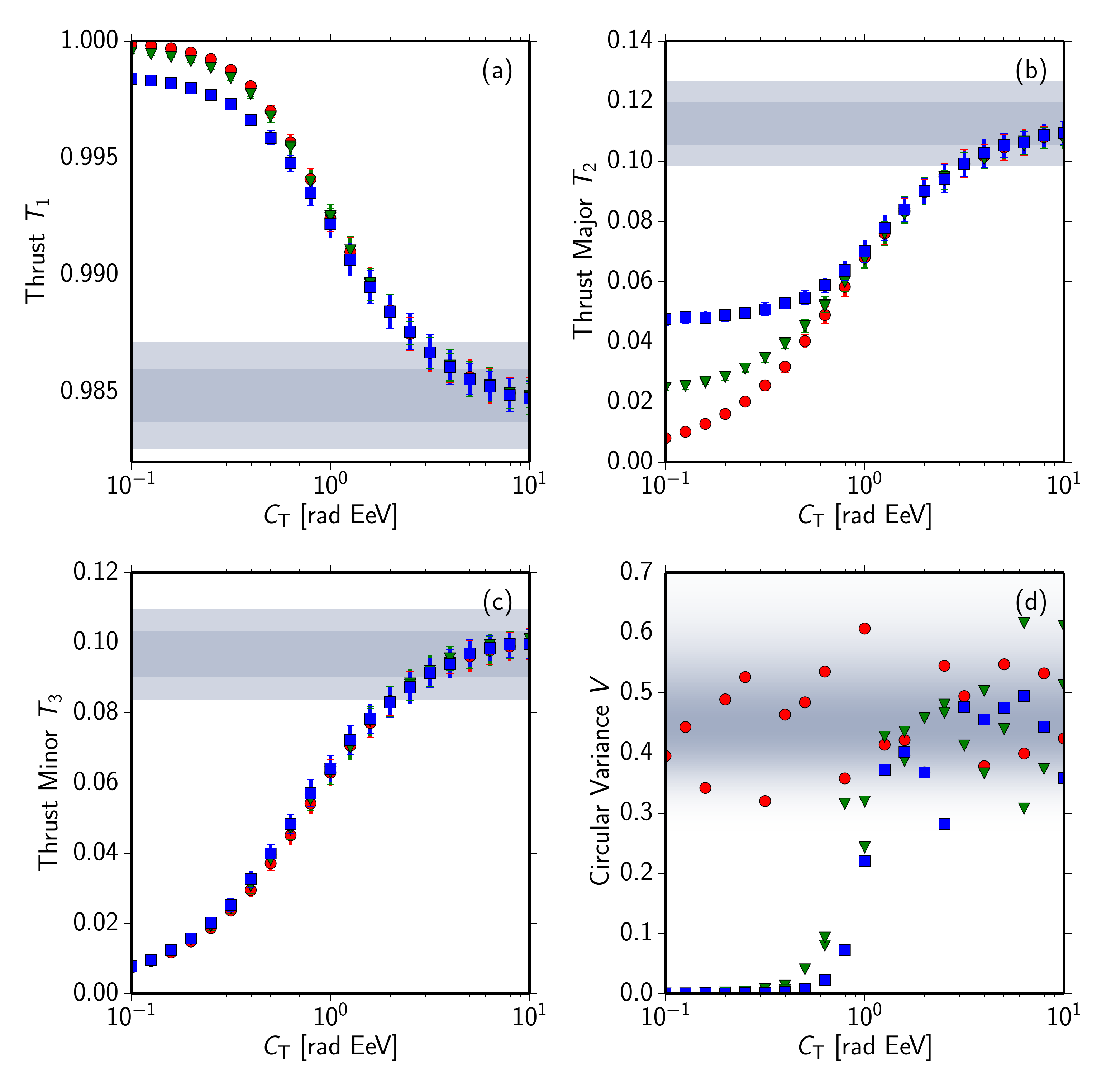}
	\caption{Response of the thrust observables to typical deflection patterns.
		\textbf{(a-c)} Mean and spread of the observables $T_{1,2,3}$ as a
		function of the strength of the deflection in turbulent magnetic
		fields $C_\text{T}$.  Red circles correspond to no directed deflection,
		green triangles to $C_\text{C} = \SI{0.5}{\radian\EeV}$ and blue squares to
		$C_\text{C} = \SI{1.0}{\radian\EeV}$. The shaded area corresponds to the
		$1\sigma$ and $2\sigma$ expectations of the observables for an
		isotropic distribution of cosmic rays.  \textbf{(d)} Circular
		variance of the thrust-major axes calculated in the simulations in 100
		ROIs. Gray shading corresponds to the probability density of the
		expectation value of the circular variance of uniformly-distributed
		directions.}
	\label{fig:ThrustObservables_ToyMC}
\end{figure*}
In Figure~\ref{fig:ThrustObservables_ToyMC}~(a-c)
the mean and spread of the
thrust observables $T_{1,2,3}$ 
of 100 realizations of the ROI at each point in the explored parameter space are shown. We used $C_\text{T}
=$\SIrange{0.1}{10}{\radian\EeV}, without coherent deflection, and alternatively with 
$C_\text{C}=\SI{0.5}{\radian\EeV}$ as well as $C_\text{C}=\SI{1.0}{\radian\EeV}$. 

All three observables are sensitive to a symmetric blurring of the
source. For increasing $C_\text{T}$ the
distribution of cosmic rays in the ROI becomes isotropic, and the observables
approach the corresponding expectation value.  The value of the thrust major
and thrust minor for strong patterns is here below the expectation for no
patterns, as the particles are concentrated in the center of the ROI.  The
thrust minor, Figure~\ref{fig:ThrustObservables_ToyMC}~(c), does not
depend on the strength of coherent deflection, as the width of the
blurring is determined here only by the strength of $C_\text{T}$.

When measuring a thrust-major axis of an individual ROI, we also want to
determine the stability of the axis direction. As explained in
Section~\ref{sec:Methods}, the thrust major-axis is located in the plane
tangential to a sphere around the observer, and provides a directional
characteristic on the sky.
We quantify the stability of the axis 
using the circular variance $V$ derived in the specialized statistics for
directional data~(e.g.~\cite{Mardia1972, Jammalamadaka2001}). The direction of
the thrust-major axis $\vec{n}_{2,i}$ in a region of interest $i$ is defined by the
angle $\theta_i$ between the axis and the local unit vector $\vec{e_\phi}$ in
spherical coordinates with $\theta_i \in [0 \ldots \pi)$.  

To calculate the circular variance $V$ from the $n$ observations
$\theta_i$, first the $\theta_i$ 
are transformed to
angles on the full circle by $\theta^*_i = \ell \cdot \theta_i$ with $\ell = 2$ owing to
	the symmetry of the thrust-major axis.  
With
\begin{equation}
	C = \sum_{i=1}^n \cos\theta^*_i,\qquad 	S = \sum_{i=1}^n \sin\theta^*_i
	\label{}
\end{equation}
the resultant length $R$ is defined as
\begin{equation}
	R = \sqrt{C^2 + S^2}.
	\label{ResultantLength}
\end{equation}
Based on the resultant length $R$ in Equation~\eqref{ResultantLength} the
circular variance $V$ of a sample of size $n$ is defined  
as
\begin{equation}
	V = 1 - \left(\frac{R}{n}\right)^{1/\ell^2}. 
	\label{eq:CircularVariance}
\end{equation}
In contrast to the variance in linear statistics, $V$ is limited to the
interval $[0,1]$.  The circular variance is a consistent measure
for the concentration of observations on periodic intervals with $V=0$ for
data from a single direction and $V=1$ for perfectly dispersed data.
Even in the limit $ n \ll \infty$  a value $V < 1$ is also expected
for non-directed data as perfect dispersion is unlikely in a random
sample.

To demonstrate the strength of correlation of the axes with the direction of
deflection in the simulation we use the circular variance $V$ among the
simulated sample as a measure. The resulting values for the 100 simulated
scenarios at every point of the aforementioned parameter space are shown in
Figure~\ref{fig:ThrustObservables_ToyMC}~(d).  In case of zero coherent
deflection, and also in case of strong blurring of the sources, no stable axis 
is found.   For small blurring of the sources, the variance between the directions
is zero, if there is coherent deflection.

%% file: Results.tex
\section{Measurement}
\label{sec:Results}

For the measurement of the observables we selected events above \SI{5}{\EeV}
recorded with the surface detector of the Pierre Auger Observatory up to March
19,  2013. We require that the zenith angle of the events is smaller than
\SI{60}{\degree} and that the detector stations surrounding the station with
the highest signal are active~\cite{PAO2010b}.  30,664 events are included in
the analysis; 70 fulfill the conditions  $E \ge \SI{60}{\EeV}$ and are at least
\SI{0.25}{\radian} inside the field of view of the Pierre Auger Observatory and
therefore seed an ROI. 

In order to estimate the uncertainty on the measurement, we repeatedly vary the
energy and arrival directions of all events detected with the Pierre Auger
Observatory above $E = \SI{3}{\EeV}$ and $\theta < \SI{60}{\degree}$ within
their experimental uncertainties and repeat the calculation of the observables
with the new values. The mean and spread of the resulting distributions then
serve as measured observables and their corresponding uncertainty. The energy
resolution of the surface detector is 16\%~\cite{DiGiulio2009} and the angular
resolution of the SD is better than $1^{\circ}$ for energies above
\SI{5}{\EeV}~\cite{Bonifazi2008}. The selected ROIs are kept fixed to the
original positions in all repetitions.  Because of the decreasing spectrum, the
number of events in the analysis increases as more events propagate above the
lower energy threshold than vice versa.  To keep the number of events in the
uncertainty analysis fixed, the 30,664 events with the highest energy after
variation are selected.
\begin{figure*}[tbp]
 	\includegraphics[width=\textwidth]{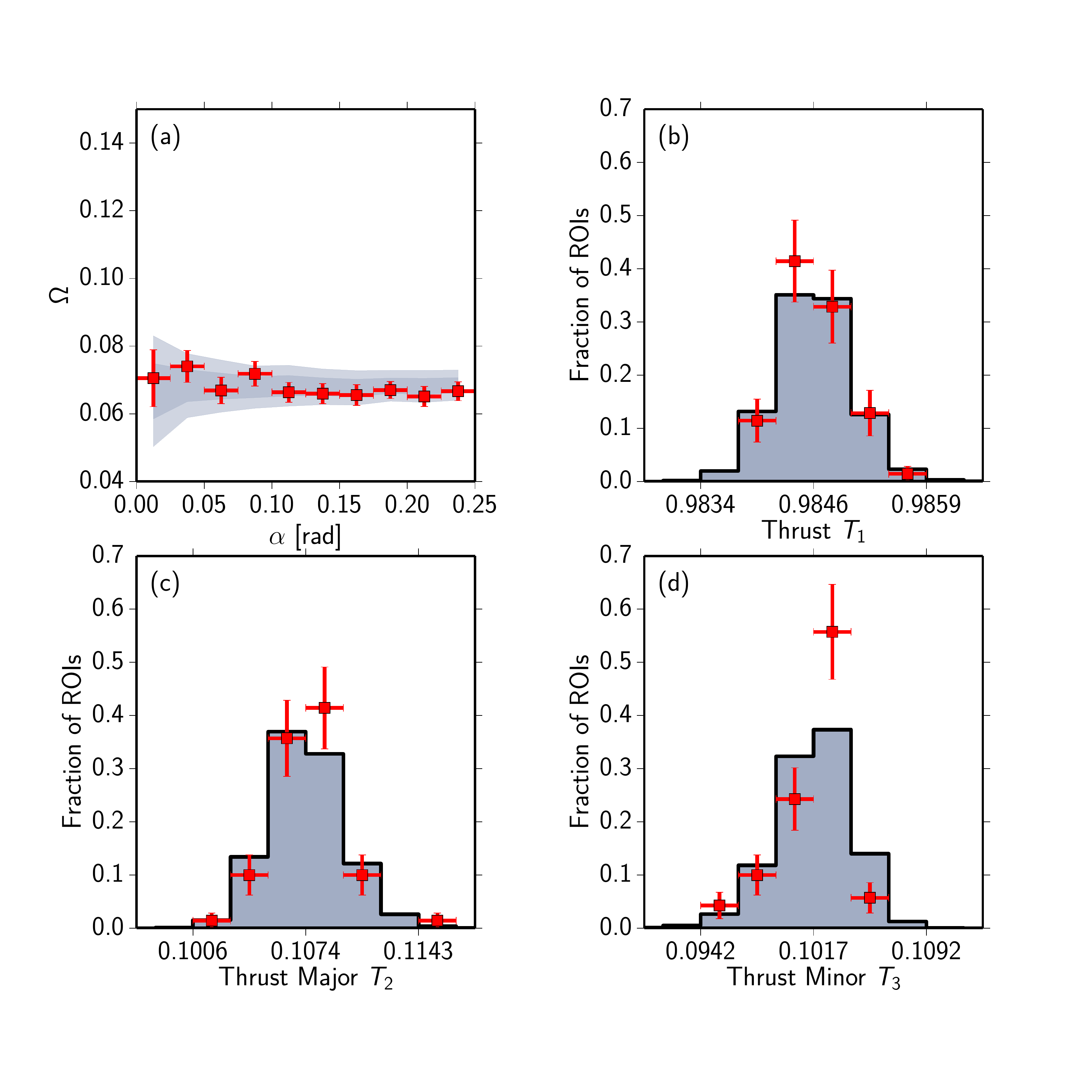}
 	\caption{Measurement of the \textbf{(a)} energy-energy correlation
 	$\Omega$ and \textbf{(b-d)} thrust observables $T_{1,2,3}$ with the
 Pierre Auger Observatory (red squares and error bars). The
 measurements are compared to distributions without structure in the arrival
 directions of UHECRs (gray distributions).}
 	\label{fig:Measurement}
 \end{figure*}

 In Figure~\ref{fig:Measurement} the distributions of the measured EEC and
 thrust observables are shown together with the distributions expected from
 isotropic arrival directions of UHECRs.  The goodness-of-fit of the
 measurements compared to expected distributions without structure in the
 arrival directions of UHECRs, using a $\chi^2$ test, yields $p$-values which are
 all above $p=0.2$ except for the thrust minor distribution with $p(T_3)=0.01$.
 Note that the $p$-value for $T_3$ results from a lack of signal-like regions
 in the data which are expected to broaden the distribution.  The measured
 distributions of all four observables reveal thus no local patterns in the
 arrival directions of UHECRs.

\begin{figure*}[htp]
		\includegraphics[width=\textwidth]{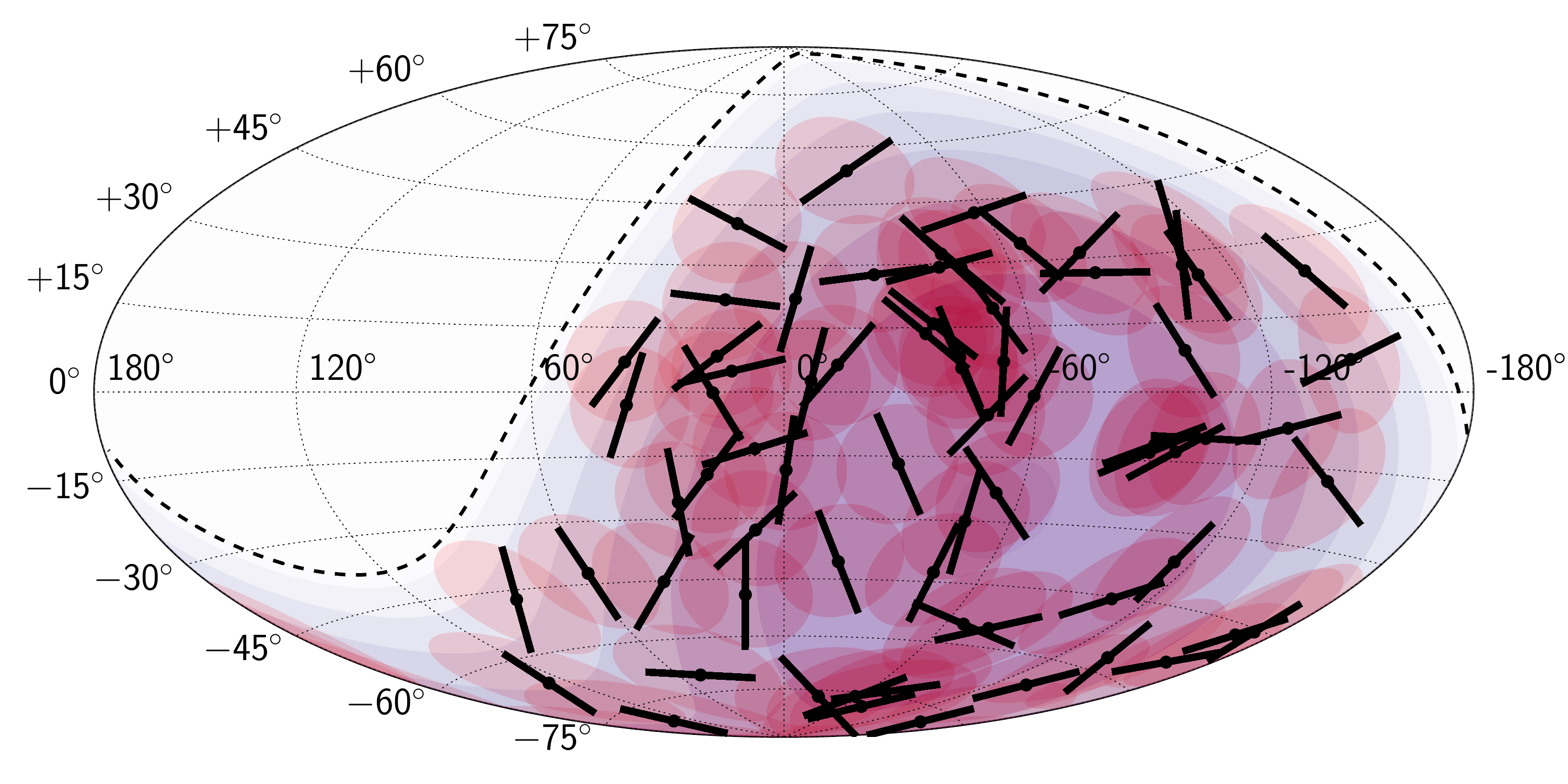}
	\caption{Hammer projection of the map of principal axes of the
		directional energy distribution in galactic coordinates. The red
		shaded areas represent the regions of interest. Black lines denote
		the second principal axes (thrust-major axes) $\vec{n}_2$, black
		dots mark the positions of the thrust axes $\vec{n}_1$. The blue
		shading indicates the exposure of the Pierre Auger Observatory; the
		dashed line marks the extent of its field of view.}\label{fig:Skymap}
\end{figure*}
From the principal-axes analysis, a map of the thrust-major axes is derived
which is shown in Figure~\ref{fig:Skymap}. If not trivial, these axes
correspond to the direction of preferred cosmic-ray deflections. 
This question is further studied in the following
section.

%% file: Discussion.tex
\section{Discussion}
\label{sec:Discussion}

In this section we first continue with analysing the directions of the thrust
axes shown as a sky map in Figure~\ref{fig:Skymap}. The aim is to search for
any individual ROI with signal contributions, e.g.~cosmic rays from a point
source, by testing the reproducibility of the axis direction. We will then
compare the measured distributions of the energy-energy correlations and the
thrust values in Figure~\ref{fig:Measurement} with astrophysical simulations
obtained with the PARSEC Monte Carlo generator. Using these comparisons, limits
on the strength of the deflection of the UHECRs in extragalactic magnetic fields
and the density of point sources of UHECRs are derived.

\subsection{Reproducibility of the Axes Measurement}
We further investigate the directional information shown by the thrust-major
axes of the individual ROIs in Figure~\ref{fig:Skymap}. From the simplified
simulations in Section~\ref{sec:ToyMC} we saw that thrust-major directions are
reproducible in repeated experiments for scenarios where coherent deflections
contribute, and turbulent deflections are not too large.  In additional
simulation studies it was shown that evidence for anisotropy could sometimes
be found in reproducibility of axis directions even when the thrust scalar
values were consistent with isotropy~\cite{Winchen2013}.  Hence, analysis of the directions of
the thrust-major axes could potentially reveal further information.

As we have obtained a single set of measured UHECR data at this point in time, we
perform here a stability test on subsets of the data in the following sense. If
the measured thrust-major direction obtained in a single ROI is related to a
deflection pattern reasonably constant in time then the analysis of subsets of the
measured data should also reflect this pattern. As only a fraction of the ROIs
may contain such a deflection pattern we perform tests of reproducibility
on each ROI individually.

We first define the ROIs as before using all available data. We then split the
dataset into $n$ independent subsamples and compare the directions
$\vec{n}_{2,j=1} \ldots \vec{n}_{2,j = n}$ obtained in each subsample for every
individual region of interest.  A low variability of directions in the subsets
of the data provides evidence for a non-triviality of the thrust-major axis and
consequently for an anisotropic distribution of UHECRs.

The optimal choice for the number of subsamples to split the data into is not
known a priori. On the one hand, a large number of $n$  maximizes the number of
repeated experiments. On the other hand, as the total number of UHECRs is fixed,
$n = 2$ maximizes the number of UHECRs in every subsample. We investigated the
choice of $n$ using simulations of the simplified model described in
Section~\ref{sec:ToyMC}. The test power to distinguish regions of interest
containing 600 anisotropically distributed UHECRs from regions with
isotropically  distributed UHECRs using the circular variance $V$ reaches a
plateau for $n \gtrsim 12$.

The dependence of the results and their variance with random splits of the data
set into 12 parts was investigated.  The observed axis directions shown in
Figure~\ref{fig:Skymap} were not reproducible in subsets of the data with this
analysis. No evidence for a non-triviality of the axes was thus found.

\subsection{Limits on Propagation Parameters}
\begin{figure*}[tbp]
	\includegraphics[width=\textwidth]{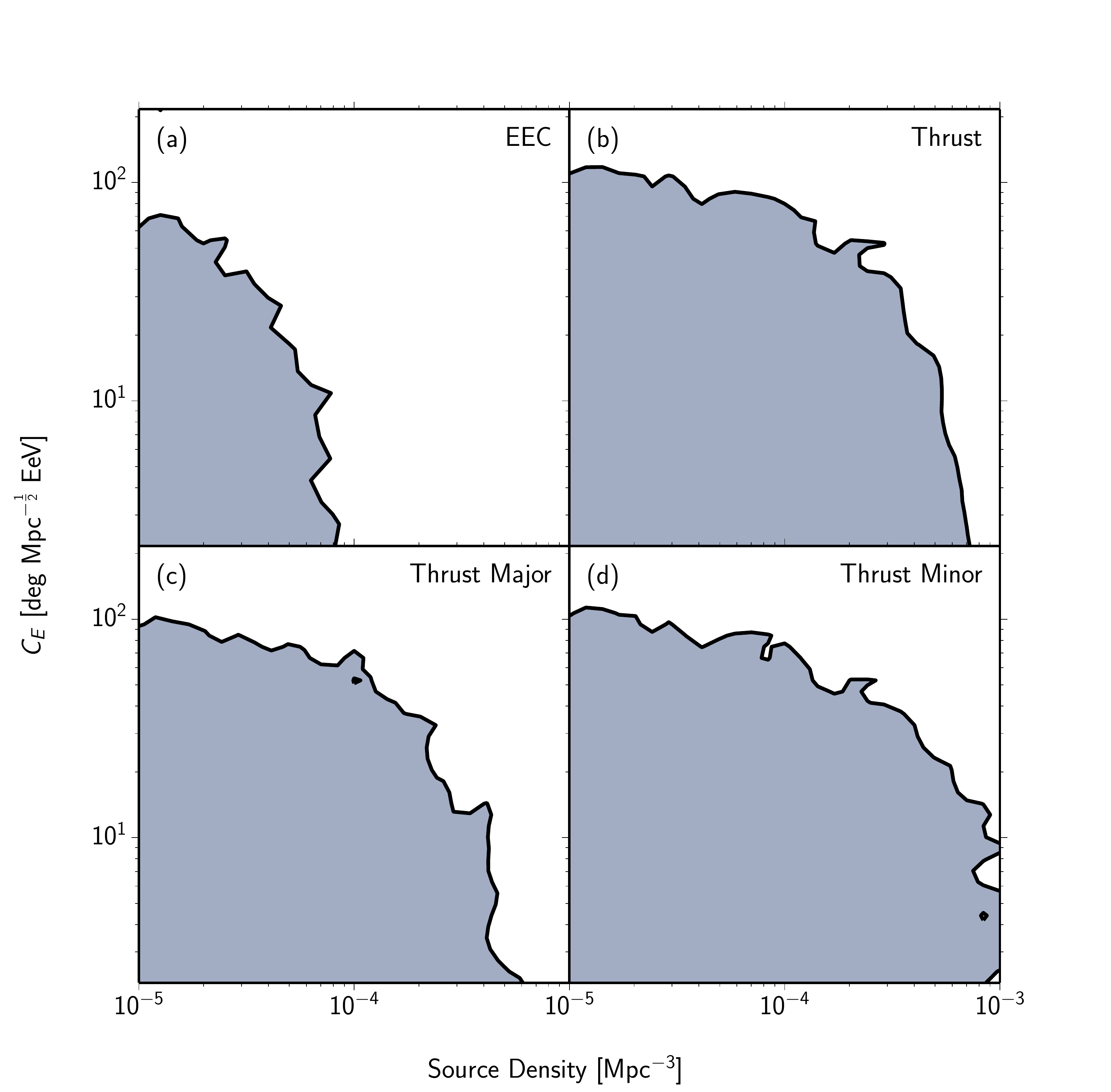}
	\caption{95\% $CL_S$ limits on the strength of the deflection of cosmic-ray protons $C_\text{E}$ (cf.~Equation~\eqref{eq:CECTRelation} and \eqref{eq:CTurbulent}~ff.) and density of point sources $\rho$ in simulations using the
	PARSEC software~\cite{Bretz2013} from the analysis of the \textbf{(a)} energy-energy correlations, \textbf{(b)} thrust, \textbf{(c)} thrust-major and \textbf{(d)} thrust-minor distributions. The gray areas 
are excluded by the measurements.}
	\label{fig:Limits}
\end{figure*}

A prime value of the measurements lies in their ability to constrain UHECR
propagation scenarios. We outline the
procedure to derive limits on scenario parameters using a simple model for
extragalactic propagation of protons based on parameterizations as implemented
in version 1.2 of the PARSEC software~\cite{Bretz2013}.  Although this model is
likely too coarse to allow definite conclusions on the sources of UHECRs, it
includes at least qualitatively the effects influencing patterns in the UHECR
distributions.  Its fast computability allows a scan of a large range of
parameter combinations in the source density and the strength of the deflection
in the extragalactic magnetic field, thus limiting these important
parameters within this model.  The procedure to obtain limits from the
measurements reported in this paper as outlined here can be applied to any other
model. 

The PARSEC software simulates ultra-high energy protons by calculating the
probability-density function (pdf) to observe a cosmic ray for discrete
directions and energies using parameterizations for energy losses and
energy-dependent deflections. In the calculations, energy losses of the UHECRs
from interaction with  extragalactic-photon backgrounds, effects from the
expansion of the universe and deflection in extragalactic magnetic fields are
accounted for using parameterizations. To account for deflections in the
galactic magnetic field, the calculated pdf is transformed using matrices
derived from backtracked UHECRs using the CRT software~\cite{Sutherland2010}.

As model for the galactic magnetic field, we use here the model proposed by
Jansson and Farrar~\cite{Jansson2012,Jansson2013}. For the random field we
assume Kolmogorov turbulences with a coherence length $L_\text{c} =
\SI{60}{\parsec}$ and a maximum wavelength $L_\text{max} \simeq
\SI{260}{\parsec}$.  We use only one realization of the random component of the
model in all simulations. The directions in the simulations are discretized
into 49,152 equal-area pixels following the HEALPix layout~\cite{Gorski2005}.
The energy is discretized into 100 log-linear spaced bins ranging from
$10^{18.5}$~eV to $10^{20.5}$~eV. Both choices result in angular and energy
bins smaller than the corresponding measurement errors.

We simulated scenarios with unstructured  point sources with density $\rho$ and
strength of the deflection of the cosmic rays 
\begin{equation}
	C_\text{T} = C_\text{E} \sqrt{D}
	\label{eq:CECTRelation}
\end{equation}
 with distance $D$ of the source.
 We scanned the parameter range $C_\text{E} =
\SIrange{2}{200}{\degree\per\mega\parsec\tothe{1/2}\EeV}$ and source densities up to $\rho =
\SI{1d-3}{\per\cubic\mega\parsec}$. We considered contributions
from sources up to a distance $D_{\max} = \SI{2}{\giga\parsec}$.  At every point
of the parameter space we simulated sets of 200 pseudo experiments with the same number of events as in the measurement presented in Section~\ref{sec:Results}. 

Since the sources of the UHECRs are randomly distributed and have a maximum
injection energy $E_{\max} = \SI{1000}{\EeV}$, some realizations do not include
sources within 43 Mpc, the maximum propagation distance 
of the most energetic particle in this
analysis. Due to the continuous energy loss approximation the maximum distance is here a hard limit and
these simulations cannot reproduce the observed energies. To restrict the reported
limits to information from the observables such scenarios are not used here.
Note that within such a scenario, the necessity of a close source could be used
as an additional constraint. The probability of including at least one source
in a pdf set can be calculated analytically (e.g.~\cite{Chandrasekhar1943}) and
is higher than $96 \%$ for source densities
greater than $\rho=\SI{1d-5}{\per\cubic\mega\parsec}$. Using this argument
alone, source densities with $\rho < \SI{1d-7}{\per\cubic\mega\parsec}$ may be
disfavored. However, the inclusion of this argument only marginally modifies
the reported limits.

Limits on the strength of the deflection and the density of point
sources in the simulation are set using the $CL_S$
method~\cite{Read2000,Read2002}. Here, 
\begin{equation}
	Q = -2 \log \frac{\mathcal{L}_\text{a}}{\mathcal{L}_0}
	\label{ieq:Likelihoodratio}
\end{equation}
is the ratio of the likelihood $\mathcal{L}_0$ of the data given isotropically distributed
UHECRs, and the likelihood $\mathcal{L}_\text{a}$ of the data given the alternative
hypothesis simulated with PARSEC. In the $CL_S$ method, not $Q$
directly, but the modified likelihood ratio 
\begin{equation}
	CL_S = \frac{P_\text{a}(Q \geq Q_\text{obs})}{1-
		P_0(Q\leq Q_\text{obs})}
	\label{eq:CLSMethod}
\end{equation}
is used as test statistic. 
Here $P_\text{a}(Q \geq Q_\text{obs})$ is the frequency with which likelihood ratios $Q$
larger than the observed value are obtained in simulations 
of the alternative hypothesis and $1-	P_0(Q\leq Q_\text{obs})$ the corresponding frequency
in simulations of the null hypothesis. 
 Points in parameter space with $CL_S < 0.05$ are excluded at
the 95\% confidence level.
The resulting limits are shown in Figure~\ref{fig:Limits} for the individual
observables.  

A combination of the limits is not attempted here as it depends on
scenario-specific correlations between the observables.  If the cosmic rays are
not protons but heavier nuclei the limits are reduced accordingly.  
For the extreme case that all cosmic rays are iron nuclei with $Z=26$ the limits
shift down by more than one order of magnitude.
For the proton
case shown in Figure~\ref{fig:Limits} the extragalactic deflection of cosmic
rays needs to be larger than
$C_\text{E} = \SIrange{10}{120}{\degree\per\mega\parsec\tothe{1/2}\EeV}$ for source
densities smaller than \SI{d-3}{\per\cubic\mega\parsec} and assuming deflections
in the galactic magnetic field as expected from the Jansson-Farrar 2012 model
with a coherence length set to $L_\text{c} =
\SI{60}{\parsec}$. The exact value depends on the source density. Without
galactic random field the limits are only marginally more constraining, choosing a
higher coherence length lowers the limits according to the stronger
deflections.

Previously, we derived from two-point correlations of UHECRs with an energy
$E>\SI{60}{\EeV}$ lower bounds on the density of uniformly distributed sources
of, e.g., $2 \times 10^{-4}\,\text{Mpc}^{-3}$ if the deflection of cosmic rays
above 60 EeV is 5 degrees~\cite{Abreu2013a}. Only the total deflection due to the
EGMF and GMF was taken into account, and no explicit model for the Galactic
magnetic field was used. An approximate comparison with the current analysis
can be performed assuming the average deflections in the EGMF and GMF add up
linearly. The average deflection of 60 EeV cosmic rays in the JF2012 field
accounts to 5 degrees. The above density therefore gives a lower limit for
negligible deflections in the EGMF.

With the current analysis we obtain for the lowest EGMF considered a limit of
$9 \times 10^{-4}\,\text{Mpc}^{-3}$ from an analysis of the Thrust Minor.  We
therefore extend the lower bound on the density of uniformly distributed
sources by a factor of more than four in the case of small 
extragalactic deflections.

%% file: Conclusions.tex
\section{Conclusions}
In this work, we characterized the distribution of UHECRs with $E >
\SI{5}{\EeV}$ in regions of \SI{0.25}{\radian} around events with $E >
\SI{60}{\EeV}$ using observables sensitive to patterns characteristic for
deflections in cosmic magnetic fields.  No such patterns have been found within
this analysis. We demonstrated the usage of this non-observation to constrain
propagation scenarios using a scenario based on parametrizations for the
propagation of UHECR protons as an example. 

Within the simulated scenario, we estimate that the strength of the deflection
in the extragalactic magnetic field has to be larger than $C_\text{E} =
\SIrange{10}{120}{\degree\per\mega\parsec\tothe{1/2}\EeV}$ for source densities
smaller than \SI{d-3}{\per\cubic\mega\parsec} assuming protons and deflections
expected from the Jansson-Farrar 2012 model for the galactic magnetic field. For protons
with an energy $ E = \SI{10}{\EeV}$ from a source at \SI{16}{\mega\parsec} this
translates to a required strength of the deflection in extragalactic space of
more than \SI{4}{\degree} if the source density is smaller than
\SI{d-3}{\per\cubic\mega\parsec} and more than \SI{25}{\degree} if the source density is smaller than
\SI{d-4}{\per\cubic\mega\parsec}.

%% file: acknowledgments.tex
\section*{Acknowledgments}

The successful installation, commissioning, and operation of the Pierre Auger Observatory would not have been possible without the strong commitment and effort from the technical and administrative staff in Malarg\"{u}e. 

We are very grateful to the following agencies and organizations for financial support: 
Comisi\'{o}n Nacional de Energ\'{\i}a At\'{o}mica, Fundaci\'{o}n Antorchas, Gobierno De La Provincia de Mendoza, Municipalidad de Malarg\"{u}e, NDM Holdings and Valle Las Le\~{n}as, in gratitude for their continuing cooperation over land access, Argentina; the Australian Research Council; Conselho Nacional de Desenvolvimento Cient\'{\i}fico e Tecnol\'{o}gico (CNPq), Financiadora de Estudos e Projetos (FINEP), Funda\c{c}\~{a}o de Amparo \`{a} Pesquisa do Estado de Rio de Janeiro (FAPERJ), S\~{a}o Paulo Research Foundation (FAPESP) Grants \# 2010/07359-6, \# 1999/05404-3, Minist\'{e}rio de Ci\^{e}ncia e Tecnologia (MCT), Brazil; MSMT-CR LG13007, 7AMB14AR005, CZ.1.05/2.1.00/03.0058 and the Czech Science Foundation grant 14-17501S, Czech Republic;  Centre de Calcul IN2P3/CNRS, Centre National de la Recherche Scientifique (CNRS), Conseil R\'{e}gional Ile-de-France, D\'{e}partement Physique Nucl\'{e}aire et Corpusculaire (PNC-IN2P3/CNRS), D\'{e}partement Sciences de l'Univers (SDU-INSU/CNRS), Institut Lagrange de Paris, ILP LABEX ANR-10-LABX-63, within the Investissements d'Avenir Programme  ANR-11-IDEX-0004-02, France; Bundesministerium f\"{u}r Bildung und Forschung (BMBF), Deutsche Forschungsgemeinschaft (DFG), Finanzministerium Baden-W\"{u}rttemberg, Helmholtz-Gemeinschaft Deutscher Forschungszentren (HGF), Ministerium f\"{u}r Wissenschaft und Forschung, Nordrhein Westfalen, Ministerium f\"{u}r Wissenschaft, Forschung und Kunst, Baden-W\"{u}rttemberg, Germany; Istituto Nazionale di Fisica Nucleare (INFN), Ministero dell'Istruzione, dell'Universit\`{a} e della Ricerca (MIUR), Gran Sasso Center for Astroparticle Physics (CFA), CETEMPS Center of Excellence, Italy; Consejo Nacional de Ciencia y Tecnolog\'{\i}a (CONACYT), Mexico; Ministerie van Onderwijs, Cultuur en Wetenschap, Nederlandse Organisatie voor Wetenschappelijk Onderzoek (NWO), Stichting voor Fundamenteel Onderzoek der Materie (FOM), Netherlands; National Centre for Research and Development, Grant Nos.ERA-NET-ASPERA/01/11 and ERA-NET-ASPERA/02/11, National Science Centre, Grant Nos. 2013/08/M/ST9/00322, 2013/08/M/ST9/00728 and HARMONIA 5 - 2013/10/M/ST9/00062, Poland; Portuguese national funds and FEDER funds within COMPETE - Programa Operacional Factores de Competitividade through Funda\c{c}\~{a}o para a Ci\^{e}ncia e a Tecnologia, Portugal; Romanian Authority for Scientific Research ANCS, CNDI-UEFISCDI partnership projects nr.20/2012 and nr.194/2012, project nr.1/ASPERA2/2012 ERA-NET, PN-II-RU-PD-2011-3-0145-17, and PN-II-RU-PD-2011-3-0062, the Minister of National  Education, Programme for research - Space Technology and Advanced Research - STAR, project number 83/2013, Romania; Slovenian Research Agency, Slovenia; Comunidad de Madrid, FEDER funds, Ministerio de Educaci\'{o}n y Ciencia, Xunta de Galicia, European Community 7th Framework Program, Grant No. FP7-PEOPLE-2012-IEF-328826, Spain; Science and Technology Facilities Council, United Kingdom; Department of Energy, Contract No. DE-AC02-07CH11359, DE-FR02-04ER41300, DE-FG02-99ER41107 and DE-SC0011689, National Science Foundation, Grant No. 0450696, The Grainger Foundation, USA; NAFOSTED, Vietnam; Marie Curie-IRSES/EPLANET, European Particle Physics Latin American Network, European Union 7th Framework Program, Grant No. PIRSES-2009-GA-246806; and UNESCO.

%% file: author_list_latex.tex
\par\noindent
{\bf The Pierre Auger Collaboration} \\
A.~Aab$^{42}$, 
P.~Abreu$^{64}$, 
M.~Aglietta$^{53}$, 
E.J.~Ahn$^{83}$, 
I.~Al Samarai$^{29}$, 
I.F.M.~Albuquerque$^{17}$, 
I.~Allekotte$^{1}$, 
J.~Allen$^{87}$, 
P.~Allison$^{89}$, 
A.~Almela$^{11,\: 8}$
% ...

\par\noindent
$^{1}$ Centro At\'{o}mico Bariloche and Instituto Balseiro (CNEA-UNCuyo-CONICET), San 
Carlos de Bariloche, 
Argentina \\
% ...
$^{8}$ Instituto de Tecnolog\'{\i}as en Detecci\'{o}n y Astropart\'{\i}culas (CNEA, CONICET, UNSAM), 
Buenos Aires, 
Argentina \\
% ...
$^{11}$ Universidad Tecnol\'{o}gica Nacional - Facultad Regional Buenos Aires, Buenos Aires,
Argentina \\
% ...
$^{17}$ Universidade de S\~{a}o Paulo, Instituto de F\'{\i}sica, S\~{a}o Paulo, SP, 
Brazil \\
% ...
$^{29}$ Institut de Physique Nucl\'{e}aire d'Orsay (IPNO), Universit\'{e} Paris 11, CNRS-IN2P3, 
Orsay, 
France \\
% ...
$^{42}$ Universit\"{a}t Siegen, Siegen, 
Germany \\
% ...
$^{53}$ Osservatorio Astrofisico di Torino  (INAF), Universit\`{a} di Torino and Sezione INFN, 
Torino, 
Italy \\
% ...
$^{64}$ Laborat\'{o}rio de Instrumenta\c{c}\~{a}o e F\'{\i}sica Experimental de Part\'{\i}culas - LIP and  
Instituto Superior T\'{e}cnico - IST, Universidade de Lisboa - UL, 
Portugal \\
% ...
$^{83}$ Fermilab, Batavia, IL, 
USA \\
% ...
$^{87}$ New York University, New York, NY, 
USA \\
% ...
$^{89}$ Ohio State University, Columbus, OH, 
USA \\
% ...
\par\noindent
(\ddag) Deceased \\
(a) Now at Konan University \\
(b) Also at the Universidad Autonoma de Chiapas on leave of absence from Cinvestav \\
(c) Now at NYU Abu Dhabi \\
(d) Now at Unidad Profesional Interdisciplinaria de Ingenier\'{\i}a y Tecnolog\'{\i}as
Avanzadas del IPN, M\'{e}xico, D.F., M\'{e}xico \\
(e) Now at Universidad Aut\'{o}noma de Chiapas, Tuxtla Guti\'{e}rrez, Chiapas, M\'{e}xico \\